\documentclass[number,preprint,3p,onecolumn]{elsarticle}

\usepackage[T1]{fontenc}
\usepackage[utf8]{inputenc}
\usepackage{amsmath,amssymb}
\usepackage{longtable,booktabs,array}
\usepackage{graphicx}
\usepackage{calc}
\usepackage{makecell}

\usepackage{pdflscape}
\usepackage[]{natbib}
\usepackage{hyperref}
\usepackage{caption}

\providecommand{\tightlist}{%
  \setlength{\itemsep}{0pt}\setlength{\parskip}{0pt}}

\journal{Journal of Clinical Epidemiology}
\begin{document}

\begin{frontmatter}
\title{meta-pipe: An LLM-agent pipeline for end-to-end automated
systematic review and meta-analysis}
\author[1]{Hsieh-Ting Lin, MD%
\corref{cor1}%
}

\author[2]{Jiunn-Tyng Yeh, MD, PhD%
}

\affiliation[1]{organization={Koo Foundation Sun Yat-Sen Cancer
Center, Department of
Oncology},city={Taipei},country={Taiwan},countrysep={,},postcodesep={}}
\affiliation[2]{organization={Duke Institute for Health
Innovation},city={Durham,
NC},country={USA},countrysep={,},postcodesep={}}

\cortext[cor1]{Corresponding author}

\begin{abstract}
\textbf{Objective}: To describe the architecture and design rationale of
meta-pipe, an open-source large language model (LLM)-agent pipeline that
integrates the complete systematic review and meta-analysis (SR/MA)
workflow --- from literature search through statistical analysis,
manuscript generation, and quality assurance --- with mandatory human
oversight at critical decision points.

\textbf{Study Design and Setting}: We developed a 10-stage modular
pipeline integrating Claude (Anthropic; Opus 4 for reasoning, Haiku 3.5
for classification) for LLM-assisted screening and extraction, Python
(\textasciitilde3,600 lines of code) for automation, R (meta, metafor,
gemtc, netmeta) for statistical analysis, and Quarto for
manuscript rendering. Five mandatory human decision points enforce
oversight. We systematically compared meta-pipe's capabilities with
five existing SR automation tools based on published documentation as
of March 2026.

\textbf{Results}: meta-pipe offers four capabilities not available in
any single existing tool: automated manuscript generation from analysis
outputs, semi-automated GRADE assessment, overclaim detection (12
predefined patterns), and dual-paradigm network meta-analysis (Bayesian
and frequentist). Estimated API cost is \$15--30 per typical 5--10 study
review. No validation data are reported; this is a system description,
not a validation study.

\textbf{Conclusion}: End-to-end AI-assisted evidence synthesis is
architecturally feasible as an open-source tool with mandatory human
oversight. Formal validation reproducing published Cochrane reviews is
underway and essential before routine use.
\end{abstract}

\end{frontmatter}

\section{Introduction}\label{introduction}

Systematic reviews and meta-analyses (SR/MA) are resource-intensive:
traditional reviews require approximately 1,100 person-hours
\citep{allen1999, michelson2019} and 67 weeks \citep{borah2017}, with
the most efficient reported workflow requiring 61 person-hours
\citep{clark2020}.

Several automation tools address this burden. Covidence extends
screening to extraction \citep{kellermeyer2018}; TrialMind combines
search, screening, and extraction \citep{wang2025trialmind}; otto-SR
automates screening through meta-analysis \citep{cao2025ottosr}; and
Nested Knowledge covers SR through network meta-analysis
\citep{kallmes2025}. A scoping review of
37 studies found LLM approaches covered 10 of 13 SR steps but concluded
they are ``not yet ready for standalone use'' \citep{lieberum2025}. The
three uncovered steps were protocol development, certainty of evidence
assessment, and interpretation --- the latter two of which meta-pipe
addresses.

Despite this progress, no open-source tool integrates statistical
analysis, manuscript generation, and quality assurance as a unified
pipeline. To address this gap, we developed meta-pipe, an open-source
LLM-agent pipeline orchestrating the complete SR/MA workflow across 10
stages with five mandatory human decision points. As a system
description --- not a validation study --- this paper contributes: (1)
the pipeline architecture with standardized data contracts; (2)
automated manuscript generation from meta-analytic outputs; (3) a
quality assurance layer (GRADE assessment and overclaim detection); and
(4) a systematic comparison with five existing tools. Formal validation
reproducing published Cochrane reviews is underway separately.

\section{Methods}\label{methods}

\subsection{Pipeline Architecture}\label{pipeline-architecture}

meta-pipe comprises 10 sequential stages (Table~\ref{tbl-stages}), each
implemented as a self-contained skill module with defined inputs,
outputs, and quality thresholds. Each skill module is a structured
prompt document containing workflow guidance, command references,
validation criteria, and input/output specifications (Supplementary S2).
Inter-stage data contracts specify file schemas at each boundary: BibTeX
for bibliography records, CSV for screening and extraction data, JSON
for configuration, and Quarto markdown for manuscript components
(Supplementary S3). Schema validation is currently documentation-based;
programmatic enforcement is planned.

Human decision points are embedded at five junctures: (1) PICO
definition and eligibility criteria; (2) screening disagreement
resolution; (3) analysis type selection (pairwise vs network
meta-analysis); (4) GRADE quality assessment; and (5) interpretation and
clinical implications. These gates retain human judgment where domain
expertise is essential, consistent with consensus that AI should augment
rather than replace human reviewers
\citep{lieberum2025, cochrane2025ai}.

\subsection{Technology Stack}\label{technology-stack}

The pipeline integrates four technologies: (a) Claude
\citep{anthropic2025claude} (Anthropic) for LLM orchestration --- using
Opus 4 with extended thinking for complex reasoning tasks (screening
adjudication, risk-of-bias assessment, manuscript generation) and
Haiku 3.5 for high-throughput simple tasks (title/abstract classification,
data validation) --- introducing a commercial API dependency that affects
reproducibility, cost, and availability; portability to other LLMs
(GPT-4, Gemini, Llama) has not been tested; prompt versioning is
partially implemented; (b) Python (\textasciitilde3,600 lines of code)
for data processing, LLM-assisted screening and extraction, validation,
and quality assurance, managed via uv with lock files; (c) R for
statistical analysis: meta v8.0 \citep{balduzzi2019}, metafor v4.8
\citep{viechtbauer2010}, gemtc v1.0 \citep{gemtc2023}, and netmeta v3.4
\citep{balduzzi2023netmeta}, with package versions locked via renv; and
(d) Quarto for manuscript rendering to HTML, DOCX, and PDF with
validated cross-references and citations.

\subsection{Search and Screening (Stages
01--03)}\label{search-and-screening-stages-0103}

Literature searches use PubMed and Scopus by default, with optional
extension to Embase and the Cochrane Library. The default selection does
not include CENTRAL or Embase, standard requirements for
Cochrane-compliant reviews; this is acknowledged as a limitation. Search
strategies are LLM-drafted based on the PICO definition and require
mandatory expert review and refinement before execution.

Title and abstract screening employs a test-retest reliability approach:
the LLM performs two independent screening passes at temperature 0.3,
and agreement is quantified using Cohen's kappa \citep{landis1977}. We
use ``test-retest reliability'' rather than ``inter-rater agreement''
because both passes use the same model with the same training data; this
measures screening decision stability under stochastic decoding
variation, not independent assessment quality. The minimum kappa
threshold is 0.60 (moderate agreement per Landis and Koch), below
Cochrane's 0.80 for dual review. Because test-retest kappa measures
intra-model stability rather than inter-rater reliability, the 0.80
benchmark is not directly applicable: kappa below 0.60 at temperature
0.3 indicates prompt ambiguity or insufficiently specific PICO criteria
requiring human review. This approach does not substitute for
independent dual review as recommended by PRISMA 2020.

\subsection{Data Extraction and Risk of Bias (Stages
04--05)}\label{data-extraction-and-risk-of-bias-stages-0405}

Full-text retrieval uses Unpaywall API integration with manual fallback.
PDF documents are processed by passing full text to the LLM context
window; the pipeline cannot reliably extract from figures, supplementary
materials, or complex multi-page tables. Data extraction employs
structured templates with automated validation (range checks,
consistency verification, confidence scoring). Confidence scores are
LLM-generated and uncalibrated; they serve as triage rather than
validated accuracy metrics. Specific failure modes include extracting
from wrong tables, confusing intention-to-treat and per-protocol
populations, and misidentifying primary endpoints. This deviates from
PRISMA 2020 Item 11d, which recommends at least two independent
extractors.

Risk of bias is assessed using RoB 2 (randomized trials) or ROBINS-I
(observational studies) with LLM-assisted domain-level judgment and
mandatory human review.

\subsection{Statistical Analysis (Stage
06)}\label{statistical-analysis-stage-06}

Pairwise meta-analyses use REML random-effects models with Hartung-Knapp
adjustment for confidence intervals. Heterogeneity is quantified using
I\textsuperscript{2}, Cochran's Q, and prediction intervals. Publication
bias is assessed using funnel plots, Egger's/Peters' tests, and
trim-and-fill analysis \citep{sterne2011}; the pipeline warns when k
\textless{} 10. Sensitivity analyses include leave-one-out analysis and
influence diagnostics.

Network meta-analysis uses a dual-paradigm approach: Bayesian (gemtc
\citep{gemtc2023} with MCMC) and frequentist (netmeta
\citep{balduzzi2023netmeta}). Discordance --- defined as
reversed top-three treatment rankings or point estimates differing by
\textgreater0.2 SMD or odds ratio ratio \textgreater1.5 --- triggers
closer examination. These are pragmatic author-chosen thresholds
informed by conventional effect size benchmarks, not formally
calibrated.

\subsection{Manuscript Generation and Quality Assurance (Stages
07--09)}\label{manuscript-generation-and-quality-assurance-stages-0709}

Manuscript assembly uses Quarto templates for each IMRaD (Introduction,
Methods, Results, and Discussion) section, reading effect estimates
directly from R output files to prevent hallucination of statistical
results. This capability is unique among current SR automation tools.

Quality assurance comprises PRISMA 2020 checklist validation (27 items),
semi-automated GRADE assessment (with mandatory human review), and
automated overclaim detection scanning for 12 predefined patterns of
unsupported claims (Supplementary S4). The pipeline implements
stage-level checkpointing for re-execution from any failure point; API
failures trigger exponential-backoff retry for up to three attempts.

\section{Results}\label{results}

\subsection{Pipeline Implementation}\label{pipeline-implementation}

The pipeline comprises 12 skill modules (10 primary stage modules plus 2
utility modules for project setup and configuration), 24 Python
automation scripts (\textasciitilde3,600 lines of code), 5 R analysis
scripts, and 74 methodology reference documents. The source code is
available under the MIT license (repository URL and Zenodo DOI to be
added upon acceptance).

\subsection{Unique Capabilities}\label{unique-capabilities}

Four capabilities distinguish meta-pipe from existing tools
(Table~\ref{tbl-comparison}). First, \emph{manuscript generation}:
meta-pipe is the only SR automation tool producing structured IMRaD
manuscripts directly from analysis outputs, with effect estimates read
from R files to prevent hallucination. Second, \emph{GRADE assessment}:
semi-automated domain-level suggestions (risk of bias, inconsistency,
indirectness, imprecision, publication bias) requiring mandatory human
adjudication; no other tool offers this. Third, \emph{overclaim
detection}: automated scanning for 12 predefined patterns of unsupported
claims, though precision and recall have not been formally evaluated.
Fourth, \emph{dual-paradigm NMA}: Bayesian (gemtc) and frequentist
(netmeta) concordance checks, tested with synthetic data but not yet
exercised on a real clinical review.

\subsection{Feature Comparison}\label{feature-comparison}

Table~\ref{tbl-comparison} compares meta-pipe with five existing SR/MA
tools based on published documentation as of March 2026. Each tool
excels within its scope; meta-pipe's positioning is best understood
through its unique contributions --- open-source end-to-end coverage,
manuscript generation, quality assurance, and GRADE --- rather than
breadth claims. We do not claim superiority in screening or extraction
accuracy, as we have no component-level validation data. otto-SR's
reproduction of 12 Cochrane reviews \citep{cao2025ottosr} represents
a validation standard that meta-pipe has not yet achieved.

\subsection{Cost Estimate}\label{cost-estimate}

API costs are estimated at \$15--30 per typical 5--10 study pairwise
meta-analysis. These are theoretical projections based on Claude API
token pricing, not empirical measurements. The approximate per-stage
breakdown is: screening \textasciitilde\$2--5, extraction
\textasciitilde\$8--18 (driven by full-text PDF processing),
risk-of-bias \textasciitilde\$2--4, and manuscript generation
\textasciitilde\$1--3. Statistical analysis incurs minimal LLM costs.
Cost scaling beyond 50 studies has not been characterized; per-stage
token logging is planned for the validation study.

\section{Discussion}\label{discussion}

\subsection{Principal Findings}\label{principal-findings}

This paper describes the architecture and design rationale of meta-pipe,
an open-source pipeline covering all 10 SR/MA stages. The contribution
is architectural, not empirical: we have not demonstrated that the
pipeline produces accurate outputs; that evidence must come from the
planned Cochrane reproduction study. What we have shown is that
end-to-end AI-assisted evidence synthesis is technically feasible as an
open-source tool with mandatory human oversight.

\subsection{Comparison with Existing
Tools}\label{comparison-with-existing-tools}

meta-pipe complements rather than competes with existing tools
(Table~\ref{tbl-comparison}). Its core differentiators --- open-source
end-to-end scope, manuscript generation, GRADE assessment, and overclaim
detection --- address capabilities not available in any single existing
tool. Users can achieve similar coverage by chaining tools (e.g.,
Covidence for screening, R for analysis, manual manuscript writing), but
meta-pipe's value lies in integrated data flow and quality gates that
reduce manual transfer steps.

Concurrent with our work, Cao et al.~presented otto-SR, an agentic LLM
pipeline that reproduced 12 Cochrane reviews with 96.7\% screening
sensitivity and 93.1\% extraction accuracy \citep{cao2025ottosr}. Where
otto-SR optimizes for full automation, meta-pipe targets de novo
synthesis with human decision points, additionally covering NMA,
manuscript generation, and GRADE. The critical difference is
evidentiary: otto-SR has demonstrated that its pipeline works; meta-pipe
has described an architecture that could work. Several caveats apply:
otto-SR is a preprint using LLM-as-judge evaluation and does not report
cost data; however, its quantitative concordance metrics represent a
validation standard meta-pipe has not achieved.

\subsection{Human Oversight and
Reproducibility}\label{human-oversight-and-reproducibility}

The five mandatory human gates align with the Cochrane 2025 AI position
statement \citep{cochrane2025ai} and developing RAISE guidelines
\citep{raise2025}. We acknowledge a tension: otto-SR's results suggest
full automation can outperform humans on individual tasks. The human
gates are thus better understood as trust-building mechanisms for novel
research questions --- where no gold-standard review exists --- rather
than proven quality improvements. Whether human oversight improves or
degrades overall pipeline accuracy is an empirical question for the
validation study.

Reproducibility is supported by renv (R) and uv (Python) lock files, but
LLM outputs are inherently stochastic and model deprecation by
commercial providers remains unresolved. The ``open source'' claim
warrants qualification: pipeline code is public, but the Claude API
backbone is proprietary; portability to alternative LLMs has not been
tested.

\subsection{Error Propagation}\label{error-propagation}

In a 10-stage sequential pipeline, errors compound: a screening false
negative eliminates studies from all downstream analyses. Under na\"{i}ve
independence assumptions, 95\% per-stage accuracy yields only 60\%
end-to-end --- though this substantially overstates the problem because
human decision points correct errors, many errors are non-compounding (a
misclassified study is simply absent rather than introducing correlated
downstream errors), and R-based statistical stages do not introduce LLM
errors. Formal error propagation analysis --- empirical or theoretical
--- is needed to characterize reliability.

\subsection{Limitations}\label{limitations}

\emph{Methodological}: no validation data (the essential next step); LLM
screening is test-retest, not PRISMA 2020 dual review; extraction
confidence scores are uncalibrated; no systematic failure mode
documentation. \emph{Technical}: commercial API dependency affecting
cost, privacy, and reproducibility; partially implemented prompt
versioning; untested LLM portability; NMA and overclaim detection not
validated on real clinical reviews. \emph{Scope}: programming skill
barrier (R, Python, command line); English-language literature only;
default databases (PubMed, Scopus) exclude CENTRAL and Embase, falling
short of Cochrane search requirements.

\subsection{Conclusions}\label{conclusions}

meta-pipe demonstrates that open-source, end-to-end AI-assisted SR/MA is
architecturally feasible when designed with mandatory human oversight
and automated quality gates. Its unique capabilities --- manuscript
generation from analysis outputs, semi-automated GRADE assessment,
overclaim detection, and dual-paradigm NMA --- are not available in any
single existing tool. Formal validation reproducing published Cochrane
reviews is the essential next step. We invite independent groups to
validate meta-pipe and encourage the development of interoperability
standards that would allow users to combine the strongest components of
different SR automation tools.

\section*{Tables}\label{tables}

\subsection{Table 1. meta-pipe Pipeline Stages}\label{tbl-stages}

\begin{longtable}[]{@{}
  >{\raggedright\arraybackslash}p{(\linewidth - 10\tabcolsep) * \real{0.1094}}
  >{\raggedright\arraybackslash}p{(\linewidth - 10\tabcolsep) * \real{0.1250}}
  >{\raggedright\arraybackslash}p{(\linewidth - 10\tabcolsep) * \real{0.1719}}
  >{\raggedright\arraybackslash}p{(\linewidth - 10\tabcolsep) * \real{0.1406}}
  >{\raggedright\arraybackslash}p{(\linewidth - 10\tabcolsep) * \real{0.2031}}
  >{\raggedright\arraybackslash}p{(\linewidth - 10\tabcolsep) * \real{0.2500}}@{}}
\caption{meta-pipe pipeline stages with key tasks, outputs, automated
quality gates, and human decision points. Stage 03b is a sub-stage of
the screening phase.}\label{tbl-stages}\tabularnewline
\toprule\noalign{}
\begin{minipage}[b]{\linewidth}\raggedright
Stage
\end{minipage} & \begin{minipage}[b]{\linewidth}\raggedright
Module
\end{minipage} & \begin{minipage}[b]{\linewidth}\raggedright
Key Tasks
\end{minipage} & \begin{minipage}[b]{\linewidth}\raggedright
Outputs
\end{minipage} & \begin{minipage}[b]{\linewidth}\raggedright
Quality Gate
\end{minipage} & \begin{minipage}[b]{\linewidth}\raggedright
Human Decision
\end{minipage} \\
\midrule\noalign{}
\endfirsthead
\toprule\noalign{}
\begin{minipage}[b]{\linewidth}\raggedright
Stage
\end{minipage} & \begin{minipage}[b]{\linewidth}\raggedright
Module
\end{minipage} & \begin{minipage}[b]{\linewidth}\raggedright
Key Tasks
\end{minipage} & \begin{minipage}[b]{\linewidth}\raggedright
Outputs
\end{minipage} & \begin{minipage}[b]{\linewidth}\raggedright
Quality Gate
\end{minipage} & \begin{minipage}[b]{\linewidth}\raggedright
Human Decision
\end{minipage} \\
\midrule\noalign{}
\endhead
\bottomrule\noalign{}
\endlastfoot
00 & Topic intake & Brainstorming, feasibility assessment & PICO
definition, eligibility criteria & Completeness check & Yes: define
scope \\
01-02 & Search \& bibliography & Database search, deduplication &
dedupe.bib, search strategy & Duplicate detection & No \\
03 & Screening & LLM-assisted screening with test-retest check &
decisions.csv, kappa statistic & Kappa $\geq$ 0.60 & Yes: resolve
disagreements \\
03b & Analysis gate & Pairwise vs NMA determination &
analysis-type-decision.md & Network geometry check & Yes: confirm
analysis type \\
04 & Full-text management & PDF retrieval, Unpaywall integration &
manifest.csv, full-text PDFs & Retrieval rate & No \\
05 & Data extraction & LLM-assisted extraction, validation &
extraction.csv, data dictionary & Range/consistency checks & Yes: review
flagged items \\
06 & Statistical analysis & Meta-analysis (pairwise or NMA) & Forest
plots, effect estimates & Model convergence & No \\
07 & Manuscript & Quarto-based assembly, rendering & IMRaD manuscript
(HTML/PDF/DOCX) & Word count, structure & No \\
08 & Peer review & GRADE assessment, SoF tables & GRADE profiles,
reviewer report & Domain completeness & Yes: quality judgments \\
09 & Quality assurance & PRISMA validation, overclaim detection & QA
report & Pattern match count & No \\
\end{longtable}

\subsection{Table 2. Feature Comparison with Existing
Tools}\label{tbl-comparison}

\begin{table}[htbp]
\caption{Feature comparison of meta-pipe with existing systematic review
automation tools. \textbf{Bold AI} indicates AI-assisted automation;
plain text indicates supported but manual or using traditional methods.
Dash (-) indicates feature not available. Based on published
documentation and peer-reviewed publications as of March 2026.
\textsuperscript{a}Covidence extraction AI is partial (select fields).
\textsuperscript{b}DistillerSR provides risk-of-bias templates and forms
but not AI-assisted assessment. \textsuperscript{c}otto-SR's preprint
does not report NMA capability; absence from the preprint does not
confirm absence of the feature. \textsuperscript{d}Implemented but not
yet validated on real clinical reviews. \textsuperscript{e}Validation
study reproducing Cochrane reviews is underway.}
\label{tbl-comparison}
\centering
\small
\renewcommand{\arraystretch}{1.3}
\begin{tabular}{@{}lcccccc@{}}
\toprule
Dimension & Covidence & TrialMind & otto-SR & \makecell{Nested\\Knowledge} & DistillerSR & \textbf{meta-pipe} \\
\midrule
\makecell[l]{\textbf{Intended}\\[1pt]\textbf{scope}} & \makecell{SR\\mgmt} & \makecell{Search$\rightarrow$\\Extract} & \makecell{Screen$\rightarrow$\\Extract+RoB} & \makecell{Full\\SR/MA} & \makecell{SR\\mgmt} & \makecell{Full\\SR/MA} \\
\addlinespace
Search strategy & - & \textbf{AI} & - & \textbf{AI} & - & \textbf{AI} \\
T/A screening & \textbf{AI} & \textbf{AI} & \textbf{AI} & \textbf{AI} & \textbf{AI} & \textbf{AI} \\
\makecell[l]{Full-text\\screening} & Manual & \textbf{AI} & \textbf{AI} & Manual & Manual & \textbf{AI} \\
Data extraction & \textbf{AI}\textsuperscript{a} & \textbf{AI} & \textbf{AI} & \textbf{AI} & \textbf{AI} & \textbf{AI} \\
Risk of bias & Template & - & \textbf{AI} & Template & Template\textsuperscript{b} & \textbf{AI} \\
\addlinespace
Pairwise MA & - & - & \makecell{R\\(metafor)} & \textbf{AI} & - & \makecell{R\\(meta/metafor)} \\
Network MA & - & - & -\textsuperscript{c} & \textbf{AI} & - & \makecell{R (gemtc/\\netmeta)\textsuperscript{d}} \\
\addlinespace
\makecell[l]{Manuscript\\generation} & - & - & - & - & - & \makecell{\textbf{AI}\\(Quarto)} \\
GRADE / QA & - & - & - & - & - & Semi-auto \\
\makecell[l]{Overclaim\\detection} & - & - & - & - & - & \textbf{AI}\textsuperscript{d} \\
\addlinespace
Open source & No & Unclear & No & No & No & \textbf{Yes} \\
\makecell[l]{Validation\\evidence} & Internal & \makecell{Trial-\\ReviewBench} & \makecell{12 Cochrane\\(preprint)} & Published & Internal & None\textsuperscript{e} \\
\bottomrule
\end{tabular}
\end{table}

\section*{Figures}\label{figures}

\subsection{Figure 1. meta-pipe Pipeline
Architecture}\label{fig-pipeline}

\begin{figure}

\centering{

\includegraphics[width=\linewidth]{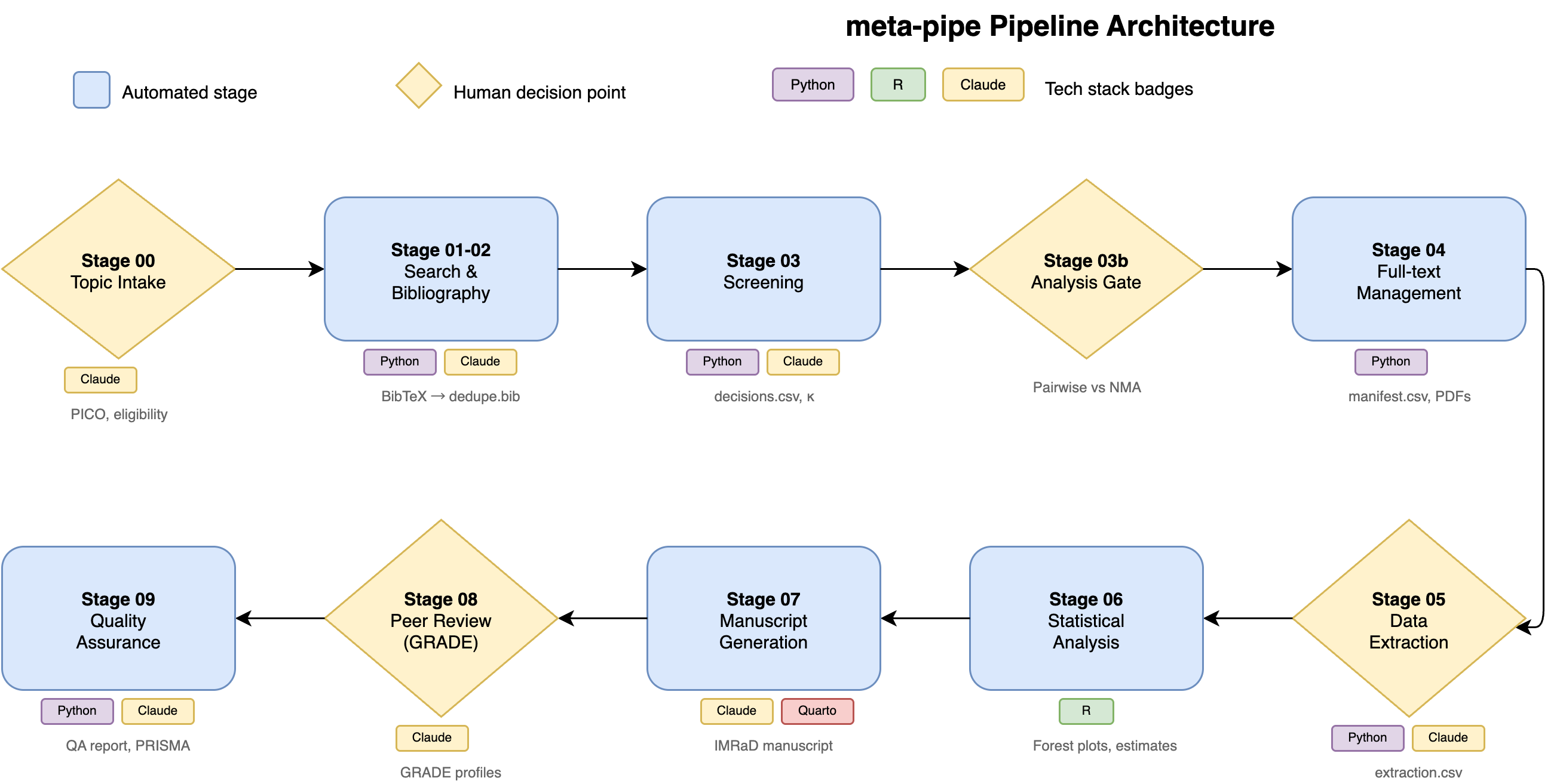}

}

\caption{\label{fig-pipeline}meta-pipe pipeline architecture. The
10-stage workflow proceeds from topic intake (Stage 00) through quality
assurance (Stage 09). Blue boxes indicate automated stages; orange
diamonds indicate mandatory human decision points. Arrows represent data
flow between stages with standardized file formats (BibTeX, CSV, JSON,
Quarto markdown) at each boundary. The pipeline integrates Claude AI
(LLM orchestration), Python (automation), R (statistical analysis), and
Quarto (manuscript rendering).}

\end{figure}%

\section*{Supplementary Materials}\label{supplementary-materials}

\subsection{S1. Representative Prompt
Templates}\label{s1.-representative-prompt-templates}

\subsubsection{S1.1 Screening Prompt
Template}\label{s1.1-screening-prompt-template}

The following is a representative prompt template used for title and
abstract screening (Stage 03). The actual prompt is parameterized with
the PICO definition and eligibility criteria defined in Stage 00.

\begin{verbatim}
You are a systematic review screening assistant. Your task is to evaluate
whether a study meets the inclusion criteria for a systematic review.

**Inclusion Criteria:**
- Population: {population}
- Intervention: {intervention}
- Comparator: {comparator}
- Outcomes: {outcomes}
- Study design: {study_design}
- Date range: {date_range}

**Instructions:**
1. Read the title and abstract below.
2. Classify as INCLUDE, EXCLUDE, or UNCERTAIN.
3. Provide a brief rationale (1-2 sentences).
4. Rate your confidence: HIGH, MODERATE, or LOW.

If the abstract lacks sufficient information to determine eligibility,
classify as UNCERTAIN (not EXCLUDE).

**Title:** {title}
**Abstract:** {abstract}

**Response format (JSON):**
{{"decision": "INCLUDE|EXCLUDE|UNCERTAIN",
  "rationale": "...",
  "confidence": "HIGH|MODERATE|LOW"}}
\end{verbatim}

\subsubsection{S1.2 Data Extraction Prompt
Template}\label{s1.2-data-extraction-prompt-template}

\begin{verbatim}
You are a data extraction assistant for systematic reviews. Extract the
following fields from the full-text article provided.

**Required fields:**
- study_id: First author last name + year (e.g., "Smith2024")
- study_design: RCT, cohort, case-control, cross-sectional, etc.
- n_total: Total number of participants randomized/enrolled
- n_intervention: Number in intervention group
- n_control: Number in control group
- intervention_description: Detailed intervention description
- control_description: Detailed control description
- primary_outcome: Name and definition of primary outcome
- effect_estimate: Point estimate (RR, OR, HR, MD, etc.)
- ci_lower: Lower bound of 95% confidence interval
- ci_upper: Upper bound of 95% confidence interval
- p_value: P-value if reported
- follow_up_duration: Duration and unit
- country: Study country/countries
- funding_source: Funding information

**Instructions:**
- Extract values exactly as reported in the paper.
- If a field is not reported, enter "NR" (not reported).
- For each field, rate confidence as HIGH, MODERATE, or LOW.
- Flag any values that seem inconsistent or require verification.
- Do NOT infer, calculate, or impute values not explicitly stated.

**Article text:**
{article_text}

**Response format (JSON):**
{{"study_id": "...", "confidence": "...", ...}}
\end{verbatim}

\subsection{S2. Skill Module
Structure}\label{s2.-skill-module-structure}

Each skill module is a markdown document with the following structure:

\begin{verbatim}
# Stage [N]: [Stage Name]

## Overview
[Brief description of the stage's purpose and role in the pipeline]

## Prerequisites
- Required input files and their schemas
- Required software and package versions
- Human decisions from prior stages

## Workflow Steps
1. [Step 1: description]
   - Command: `python scripts/[script_name].py --input [file] --output [file]`
   - Expected output: [description]
   - Validation: [check to perform]
2. [Step 2: description]
   ...

## Quality Gates
- [Gate 1]: [threshold and action if not met]
- [Gate 2]: [threshold and action if not met]

## Human Decision Point (if applicable)
- **Trigger**: [condition]
- **Required action**: [what the human must decide]
- **Documentation**: [how the decision is recorded]

## Output Specification
- File: [filename and format]
- Schema: [column headers or JSON structure]
- Validation: [automated checks before passing to next stage]

## Troubleshooting
- [Common issue 1]: [resolution]
- [Common issue 2]: [resolution]
\end{verbatim}

\subsection{S3. Inter-Stage Data
Contracts}\label{s3.-inter-stage-data-contracts}

\begin{longtable}[]{@{}
  >{\raggedright\arraybackslash}p{(\linewidth - 6\tabcolsep) * \real{0.2778}}
  >{\raggedright\arraybackslash}p{(\linewidth - 6\tabcolsep) * \real{0.1481}}
  >{\raggedright\arraybackslash}p{(\linewidth - 6\tabcolsep) * \real{0.3519}}
  >{\raggedright\arraybackslash}p{(\linewidth - 6\tabcolsep) * \real{0.2222}}@{}}
\toprule\noalign{}
\begin{minipage}[b]{\linewidth}\raggedright
Stage Boundary
\end{minipage} & \begin{minipage}[b]{\linewidth}\raggedright
Format
\end{minipage} & \begin{minipage}[b]{\linewidth}\raggedright
Key Fields / Schema
\end{minipage} & \begin{minipage}[b]{\linewidth}\raggedright
Validation
\end{minipage} \\
\midrule\noalign{}
\endhead
\bottomrule\noalign{}
\endlastfoot
00 $\rightarrow$ 01 & JSON & \texttt{pico.json}: population, intervention,
comparator, outcomes, study\_design, date\_range, databases &
Documentation only \\
01 $\rightarrow$ 02 & BibTeX & Standard BibTeX fields (author, title, year, journal,
abstract, doi) & Documentation only \\
02 $\rightarrow$ 03 & BibTeX & \texttt{dedupe.bib}: deduplicated bibliography with
unique IDs & Automated (duplicate check) \\
03 $\rightarrow$ 03b & CSV & \texttt{decisions.csv}: study\_id, title, decision
(include/exclude/uncertain), rationale, confidence, pass1, pass2,
agreement & Automated (kappa computation) \\
03b $\rightarrow$ 04 & Markdown & \texttt{analysis-type-decision.md}: pairwise or
NMA, rationale, human approval & Documentation only \\
04 $\rightarrow$ 05 & CSV & \texttt{manifest.csv}: study\_id, pdf\_path,
retrieval\_method (unpaywall/manual), retrieval\_status & Automated
(file existence check) \\
05 $\rightarrow$ 06 & CSV & \texttt{extraction.csv}: study\_id + all extraction
fields defined in template, with confidence ratings & Automated
(range/consistency checks) \\
06 $\rightarrow$ 07 & Multiple & R output files (.rds, .csv): effect estimates,
heterogeneity stats, forest plot data, funnel plot data & Automated
(model convergence) \\
07 $\rightarrow$ 08 & Quarto & IMRaD manuscript sections (.qmd), figures (.png),
tables (.csv) & Automated (structure/word count) \\
08 $\rightarrow$ 09 & JSON & \texttt{grade\_assessment.json}: outcome,
overall\_rating, domain ratings (rob, inconsistency, indirectness,
imprecision, pub\_bias), rationale per domain & Documentation only \\
\end{longtable}

\subsection{S4. Overclaim Detection
Patterns}\label{s4.-overclaim-detection-patterns}

The automated overclaim detection system (Stage 09) scans generated
manuscript text for the following 12 patterns:

\begin{longtable}[]{@{}
  >{\raggedright\arraybackslash}p{(\linewidth - 6\tabcolsep) * \real{0.0652}}
  >{\raggedright\arraybackslash}p{(\linewidth - 6\tabcolsep) * \real{0.1957}}
  >{\raggedright\arraybackslash}p{(\linewidth - 6\tabcolsep) * \real{0.2826}}
  >{\raggedright\arraybackslash}p{(\linewidth - 6\tabcolsep) * \real{0.4565}}@{}}
\toprule\noalign{}
\begin{minipage}[b]{\linewidth}\raggedright
\#
\end{minipage} & \begin{minipage}[b]{\linewidth}\raggedright
Pattern
\end{minipage} & \begin{minipage}[b]{\linewidth}\raggedright
Description
\end{minipage} & \begin{minipage}[b]{\linewidth}\raggedright
Example Flagged Text
\end{minipage} \\
\midrule\noalign{}
\endhead
\bottomrule\noalign{}
\endlastfoot
1 & Causal language for non-randomized data & Causal verbs (caused, led
to, resulted in) used for observational findings & ``Smoking
\emph{caused} increased mortality'' (in cohort study) \\
2 & Overgeneralization beyond study population & Claims extending beyond
the actual study population & ``All patients with cancer
should\ldots{}'' (when study was TNBC only) \\
3 & Recommendations without evidence level & Treatment recommendations
without stating evidence quality & ``Clinicians should
prescribe\ldots{}'' (without GRADE context) \\
4 & Unsupported superlatives & ``First,'' ``only,'' ``best,'' ``most
effective'' without citation & ``The \emph{most effective}
treatment\ldots{}'' \\
5 & Certainty language for uncertain findings & ``Clearly,''
``definitely,'' ``proves'' for findings with wide CIs & ``This
\emph{clearly demonstrates}\ldots{}'' \\
6 & Extrapolation to unexamined outcomes & Claims about outcomes not
measured in included studies & Safety conclusions drawn from
efficacy-only data \\
7 & Ignoring heterogeneity & Definitive statements when I\textsuperscript{2}
\textgreater{} 50\% & ``Treatment consistently improves\ldots{}'' (when
I\textsuperscript{2}=72\%) \\
8 & P-value misinterpretation & Equating statistical significance with
clinical importance & ``Highly significant (p\textless0.001)'' implying
large effect \\
9 & Absence of evidence as evidence of absence & Claiming no effect when
studies were underpowered & ``No difference in mortality was found'' (2
trials, N=200) \\
10 & Selective reporting language & Emphasizing favorable outcomes while
minimizing harms & Extensive efficacy discussion with one-sentence
safety \\
11 & Temporal extrapolation & Long-term conclusions from short follow-up
& ``Long-term survival benefit'' from 12-month data \\
12 & Comparison with absent comparator & Claims about superiority over
treatments not studied & ``Superior to standard chemotherapy'' (when not
directly compared) \\
\end{longtable}

\subsection{S5. Feature Comparison
Methodology}\label{s5.-feature-comparison-methodology}

The feature comparison in Table 2 was constructed as follows:

\begin{enumerate}
\def\labelenumi{\arabic{enumi}.}
\tightlist
\item
  \textbf{Source identification}: For each tool, we identified the
  primary peer-reviewed publication, official documentation website, and
  most recent user guides (accessed January--March 2026).
\item
  \textbf{Feature classification}: Features were classified as
  \textbf{AI-assisted} (bold) when the tool uses machine learning or LLM
  automation for that stage, \textbf{supported} (plain text) when the
  tool provides templates, forms, or manual workflows, or
  \textbf{absent} (dash) when the feature is not available.
\item
  \textbf{Verification}: Two authors independently classified features
  for each tool. Disagreements were resolved by consulting the tool's
  documentation.
\item
  \textbf{Limitations}: This comparison reflects published capabilities
  and may not capture unreleased features, beta functionality, or recent
  updates. Commercial tools may have added capabilities since our
  assessment date.
\item
  \textbf{Validation evidence}: Validation claims are based on the
  strongest published evidence (peer-reviewed \textgreater{} preprint
  \textgreater{} internal reports \textgreater{} no published evidence).
\end{enumerate}

Sources consulted:

\begin{itemize}
\tightlist
\item
  Covidence: Kellermeyer et al.~2018, JMLA; covidence.org documentation
\item
  TrialMind: Wang et al.~2025, npj Digital Medicine
\item
  otto-SR: Cao et al.~2025, medRxiv preprint; ottosr.com
\item
  Nested Knowledge: Kallmes et al.~2025, Cochrane Evidence Synthesis and
  Methods; nested-knowledge.com
\item
  DistillerSR: distillersr.com documentation (no primary peer-reviewed
  validation publication identified)
\end{itemize}

\section*{References}\label{references}

\renewcommand{\bibsection}{}
\bibliography{references.bib}

\end{document}